# Why pinning by surface irregularities can explain the peak effect in transport properties and neutron diffraction results in $NbSe_2$ and Bi-2212 crystals?

*Charles Simon[1,*], Alain Pautrat[1], Christophe Goupil[1], Joseph Scola[1], Patrice Mathieu[2], Annie Brûlet[3], Antoine Ruyter[4], M. J. Higgins[5], Shobo Battacharya[6]*

[1] laboratoire CRISMAT, CNRS - ENSICAEN, F14050 Caen,France
[2] Laboratoire Pierre Aigrain ENS Paris, France
[3]Laboratoire Léon Brillouin, Saclay, France
[4]LEMA, Tours, France
[5]NEC, Princeton, USA
[6]Tata Institute, Mumbai, India.

The existence of a peak effect in transport properties (a maximum of the critical current as function of magnetic field) is a well-known but still intriguing feature of type II superconductors such as $NbSe_2$ and Bi-2212. Using a model of pinning by surface irregularities in anisotropic superconductors, we have developed a calculation of the critical current which allows estimating quantitatively the critical current in both the high critical current phase and in the low critical current phase. The only adjustable parameter of this model is the angle of the vortices at the surface. The agreement between the measurements and the model is really very impressive. In this framework, the anomalous dynamical properties close to the peak effect is due to co-existence of two different vortex states with different critical currents. Recent neutron diffraction data in $NbSe_2$ crystals in presence of transport current support this point of view.

## Introduction

The understanding of the values of the critical current in a type II superconductor is still a challenging problem. The competition between elastic properties and disorder induced by pinning leads theoretically to different vortex matter states. In this respect, the peak effect observed in few type II superconductors, i.e. a sudden increase of the critical current close to the superconducting-normal transition, has been for a long time considered as a proof of a disorder transition in the vortex lattice. Bulk pinning centers can become more effective on a less rigid lattice at high field and that can lead to a peak effect. Larkin-Ovchinikov collective pinning model /1/ provides a more precise theoretical approach to the link between the loss of long range order and the high critical current. Numerous experiments have also shown that the link between vortex lattice order and critical current, is far from being direct, in contradiction the previous assumptions. Thorel's neutron scattering experiments have first shown that the Flux Lines Lattice (FLL) quality can be modified without changing the critical current /2/. A possible explanation, already proposed by different authors /3/, is that the exact order of the FLL in the bulk is not the most important parameter governing transport properties.

In addition to the peak effect, very peculiar transport properties are observed in the same region of the phase diagram, in particular hysteretic V(I) curves /4/. A model has recently emerged, supported by different experiments made in 2H-$NbSe_2$ /5/. The key ingredients are a supercooling of a high critical current into a low critical current state, and an annealing effect over surface barrier. To explain the high critical current phase, it is usual to involve a strongly disordered FLL, amorphous or liquid like, created after a genuine phase transition through the peak effect. Nevertheless, very little is known about the genuine bulk structure of these phases. There is even contradictory and puzzling results. Indeed, recent decoration experiments have shown that no disordered state can be specially evidenced in the peak effect region of pure or Fe doped $NbSe_2$ samples /6/. The high critical current FLL state remains unexplained.

* Email: presenting.charles.simon@ensicaen.fr

In the present paper, we will review our recent and less recent results about different methods to test the role the surface pinning in the critical current values. In a first part, we will discuss the model of pinning by surface irregularities to calculate the critical current without any adjustable parameter. Then we will discuss the case of the peak effect and conclude on to Bi-2212 system in which the very large anisotropy makes difficult to calculate the elasticity.

**Surface treatments in Nb films**

The sample used is a film of Niobium (thickness=3000Å) deposited at 780C on a sapphire substrate by the ion beam technique. The film has a resistivity of about 0.5 μΩ.cm at the critical temperature T c =9.15K and exhibits a surface rms roughness 5nm, measured by Atomic Force Microscopy (Nanoscope III,Digital Instruments). Microbridges of W=10 μm ×L=30 μm have been patterned using a scanning electronic microscope, this irradiation step being followed by a reactive ion etching process. The critical currents have been measured by mean of the standard four-probe technique, at 4.2 K in the whole range of field covering the mixed state. The critical current values Ic were determined with a voltage criteria of 10 nV. We have therefore measured the microbridge roughness using AFM in tapping mode. The next step was to use a Focused Ion Beam to etch its surface and then to modify the surface structure. Following a simple analysis of the surface roughness described elsewhere /7/, we have extracted a roughness $\alpha$ of about 2.2° for the rough surface to be compared to 0.6° before the processing. On figure 1, we have presented the critical current measured as function of magnetic field B at 4.2K. In order to analyse the data, we have also presented on the same figure the reversible magnetization $\varepsilon$. In the MS theory of surface pinning, there is a direct relation between these two quantities in the case of an isotropic type II superconductor for a single surface:

$$Ic/W = \varepsilon \sin(\alpha) \qquad (1)$$

In the inset of the upper figure 1, we have reported the value of $\alpha$ calculated from equation 1. One can see that $\alpha$ is roughly constant and equal to what is observed by AFM. The damaged surface presents a roughness which is increased from 0.8 to 1.6 degrees. One can see that it gives very reasonable values, with a rather small magnetic field dependence. This type of analysis gives a simple explanation for the high critical current density observed in this kind of clean thin films, compared to the moderated one observed in bulk crystals. It gives also evidence that the interaction between the surface corrugation and the vortex elasticity is a key point for the understanding of vortex lattice pinning and dynamics.

It is out of the scope of this paper to describe in detail all the subtleties of the MS theory that can be found elsewhere /3,8,9,10/ but let us summarize in few words the idea of this model. The two fundamental equations of this model are the Maxwell equation and the current conservation at the sample surface:

$$Js + \mathrm{curl}\ \varepsilon = 0 \qquad (2)$$

and $\varepsilon \times n = 0$.                (3)

(Js the superconducting current and n the vector perpendicular to the perfect flat surface). A very important point of the model is that $\varepsilon$ is a function not on the magnetic field B, but of the local field $\omega$ which is given by $\omega = B - 1/\mu_0$ curl Js. In the ideal case where the vortices are perpendicular to the ideal flat surface, Js = 0. In the case of a real surface, the vortices can join the surface with a maximum angle $\alpha$ and a non dissipative current may flow close to the surface ($Js=-\varepsilon(\omega) \sin(\alpha)$ in the isotropic case). Above the critical current, the vortices start to move with a velocity which is proportional to I-Ic which is the dissipative current flowing in the sample thickness. This perfectly explains the shape of the I-V curves (V=R(I-Ic) where Ic= Js w). In the case of an anisotropic sample, the angle $\alpha$ is the angle between $\varepsilon$ and the vector n, but the vortex lines $\omega$ are inclined by a larger angle $\theta$ such as $\tan\theta = \gamma^2 \tan\alpha$. For example, if $\alpha = 1°$, $\theta=9°$ for $\gamma=3$. Following the calculation detailed in reference 8, one finds

$$Ic^0 = W (B_{c2}^*-B)/( 2\mu_0 1.16\gamma^2 K^2) \qquad (4)$$

where $B_{c2}^*= B_{c2} (1+\gamma^2\tan^2\alpha)^{-1/2}$        (5)

**Surface treatments in $NbSe_2$ crystals**

The fact that this model works quite well in a thin film is indeed impressive, but the same type of results can be observed in $NbSe_2$ crystals. Large single crystals of H-$NbSe_2$ (size 8 x 6 x 0.5 $mm^3$, Tc = 7.5K measured by specific heat) were used. The magnetic field was applied parallel to the c-axis of the crystal In figure 2, we have presented some typical results in these crystals. The data can be fitted easily by the MS model with a single adjustable parameter irregularities $\alpha$ of 0.9° for the pristine sample. One can note that in the Abrikosov regime, $\varepsilon$ decreases linearly versus $\omega$. By sandblasting the surface, one increases $\alpha$ from 0.9° to 2.4°. By cleaving the sample after sanding, it decreases back to 1.1°. In each case, the critical current follows the model described in equation 4 and 5. One should note here that $B_{c2}^*$ decreases as $\alpha$ increases as it can be seen on figure 2. The three others parameters of the fit ($B_{C2}$ , $\gamma$ and K) are obtained by independent magnetic measurements. These values of surface roughness are in good agreement with AFM observation. One should also note that the critical current does not depend on the thickness, so cleaving the sample does not decrease the critical current. One can also see on figure 2 that a very important effect is not explained by the model, i.e. the peak effect close to $B_{C2}$.

In the peak effect, the critical current Ic is larger than the calculated one $Ic^0$. One can assumes that the current is also in surface and calculate the associated critical angle $\alpha$. It gives 8° whatever is the surface state. An important observation can be done on the data which supports this assumption. The magnetic field $\omega$ close to the surface is increased by the curvature of the flux lines ($\omega_{surface}=\omega_{applied}/\cos(\theta)$). For this reason, the critical current vanishes for $\alpha=8°$ before that for $\alpha=1°$ as one can see in formula (4). This is clearly observed on the data here, but also in previous publications.

In fact, the nature of this critical current is different from the previous one since the shape of the I-V curve is very different: even when the critical current is the high value, the curve extrapolates at high currents to the small critical current value. In addition, spatially resolved measurements exhibit two different phases (the high critical current close to the edge, the small one in the sample center). One can see the S-shape of the I-V curve as a continuous progression of percentage of the low critical current phase as one increases the current in the sample. $V=xR(I- Ic^0)$ where x is the phase fraction. In the following, we have summarized the process of dissipation above Ic: due to edge effects, the magnetic field in the sample center is slightly larger than close to the edges, so dissipation will appear there first. As soon as dissipation appears, a quite large dissipation appears close to the surface, leading to a damped oscillation of the vortex head which will reduce the stability of the vortex pinning at large angle. This effect will spreads out the current into the sample bulk. The presence of induced vortex loops perpendicular to the main magnetic field will act as a zip and depins the large critical current area. In fact, the situation is very similar to solid friction in which the critical force necessary to start the motion is always larger than the friction force in movement. The analogy here can be seen as the following: immobile, the vortex lattice may find large surface angles. When it starts to move, dissipation in the surface region creates additional “heating” (or breathing of the vortices) which reduces the stability of the large curvatures. Once the movement starts, it is impossible to recover the stability.

However, two questions remain:

- The “magic” angle 8° is probably related to the surface properties of $NbSe_2$ crystals. It is difficult to be more precise in the frame of this model but one can imagine that the step edges are surface defects which are not modified by the sandblasting and can be at the origin of the magic 8°.
- When the magnetic field decreases, the critical current increases and the surface dissipation IcV increases. The density of the perpendicular vortex loops also increases. It is clear that a limit should appear to the stability of the high critical angle phase. This interpretation of the data is supported by the fact that it is possible for example, by field cooling the sample, to stabilize metastable high critical current phases even below B/Bc2=0.75. The low magnetic field limit of the high critical current phase appears to be crucially related to experimental conditions and due to metastable states.

The present data suggest that the critical current in the peak effect is due to surface pinning. The metastable equilibrium of the vortices can be reached by special preparation of the vortex state. This unstable state can be destabilized in the sample center by a too large current and propagated to the whole sample by the presence of perpendicular vortex loops. Sandblasting does not modify this high critical current state so the irregularities responsible of this pinning should be at a different scale such as step edges of the cleaved surface.

## Transport data in Bi-2212 crystals

Let us now study the case of a very anisotropic sample such is Bi-2212. The samples used in this study are slightly Pb doped single crystals of the Bi-2212 family ($Bi_{1.8}Pb_{0.2}Sr_2CaCu_2O_{8-\delta}$). They were grown by the self-flux technique as previously described. Each cleaved single crystal was laser tailored in the form of a microbridge with a controlled pattern of ($W=200 * L=400\ \mu m^2$). The crystal was annealed under a controlled pure oxygen gas flow and is in the slightly overdoped regime (Tc= 79.5 K). Low resistance electrical contacts were made by bonding gold wires with silver epoxy. The DC transport measurements were performed using a standard four probe /11/.

We have performed V(I) curves at low temperature (T = 5K) in order to minimize thermal fluctuations. Let us first discuss the results for high magnetic field values. The V(I) curves present the usual linear form as soon as I is slightly higher than the critical current. There is no evidence of an ohmic regime at low applied current and the depinning is rather stiff. Furthermore, when comparing the effect of Field Cooling (FC) and Zero Field Cooling (ZFC), or FC under different cooling rates, we measure the same dissipation in the time scale of our experiment. In particular, no aging effect is observed on the critical current, what is not in agreement with a glassy nature of the VL governing transport properties.

When the magnetic field is decreased, we observe a different behavior in a restricted region of the phase diagram. When the vortex lattice is prepared after FC, the V(I) curves exhibit a S-shape with a high threshold current but only for the first ramp of current. After, Ic is always obtained (Fig. 3). This has been previously observed in the pulse current experiments, and this "high threshold current state" has been evidenced as a metastable state with a very long relaxation time /11/. Our measurement using a dc current evidences that the observation of this state is not due to the kind of stimulation used. Concerning the metastable V(I) curves, the peak effect in the critical current, the coexistence of two VL states, the same kind of behavior is currently observed in $NbSe_2$. The strong difference is that the peak effect and the associated metastable effects appear close to $B_{c2}$ in $NbSe_2$ but is here restricted to a very low field value. As the temperature applied in both experiments is similar, it is likely that the explanation of this field value difference has to be found in the large difference in the electronic anisotropy. For field lower than about 0.05T, we do not observe any hysteresis within the V(I) curves.

One has to say that the variation of the critical current, if one excepts the small low field part where metastability takes place, looks really like what is measured in soft low Tc materials. To some extend, one can speculate that the same pinning mechanism is acting without the need of a transition in the VL. Qualitatively, the functional form of the critical current is very close to that of the reversible magnetization of a high $\kappa$ anisotropic superconductor i.e. Ic is directly linked to the weight of the diamagnetic screening currents.

The case of very anisotropic samples is especially interesting, because it is predicted that for not too low magnetic field values and for realistic accessible surface roughness, the surface critical current becomes independent of the surface quality and solely depends on parameters of the condensate. For clarity, we restrict the comparison to the high field values in order to use the Abrikosov limiting expressions. One expects /8/

$$Ic/W= B_{c2}/(2\mu_0\beta\kappa)\ \ (1-B/B_{c2}{}^{2/3})^{3/2} \qquad (6)$$

One can see on the figure 4 that the agreement with the experimental data is really very good. The exact understanding of the peak effect is slightly more delicate. One can estimate what should be the $B_{c2}{}^*$ using the formula (5). If $\gamma$ is very large (typically 60 here), one can develop equation (5) neglecting 1 compared to $\gamma^2\tan^2\alpha$ into

$$B_{c2}{}^*= B_{c2} / (\gamma \tan\alpha) \qquad (7)$$

which is typically 2000G here, in very good agreement with the experimental value of fig. 4. The difficulty to give an exact value of Jc is that $\varepsilon$ (B) is not known in Bi-2212 where $\kappa$ is the order of 100, too large for any simple approximation of $\varepsilon$ if $B/B_{c2}$ is typically $10^{-6}$. This is something which should be calculated exactly.

## Small angle neutron scattering in Pb-In and $NbSe_2$ crystals

Small Angle Neutron Scattering thus appears as a unique technique, since the order of the FLL can be tested. It is also possible to measure in-situ V(I) curves together with the FLL diffraction and hence to investigate the relationship between the current distribution and the FLL structure in the sample. One constraint is the respect of the Maxwell-Ampere equation, lying both field lines density and bending with the current density /12,13/. The aim of these types of experiments is to use SANS in order to compare different FLL states with respect to their dynamical properties. We will focus on the case of FLL states close to the peak effect.

The Small angle Neutron scattering (SANS) experiments were performed in the Laboratoire Leon Brillouin (Saclay, France). Large single crystals of Fe doped H-$NbSe_2$ (200 ppm of Fe, size 8 x 6 x 0.5 $mm^3$, Tc = 5.5K) and of polycristalline Pb-In (10.5 % of In by weight, size 30 x 5.5 x 0.5 $mm^3$, Tc = 7K) were used. The magnetic field was applied parallel to the c-axis of the crystal and to the incident neutron beam. The scattered neutrons of wavelength 10Å were detected by a 2D detector located at a distance of 6.875 m. Superconducting leads were attached using Indium solder pressed between copper slabs: they gave us the possibility of passing a high enough transport current for this experiment. At the working temperature of 2 K (in condensed superfluid He), we can pass about 8 A without overheating. In the following, $\omega$ will refer to a rotation around the vertical axis, and $\phi$ to a rotation around the horizontal axis.

Before describing the results obtained in Fe doped 2H-$NbSe_2$, it can be interesting to compare to what is observed in a more conventional case (without peak effect), where anomalous dynamical properties are not observed. In Pb-In samples, the dynamical properties of FLL are well known. For a state defined by the two variables B and T, one measures only a single critical current. The diffraction patterns exhibit an ordered hexagonal lattice. We performed $\omega$ rocking curves. We obtain, for the ZFC FLL and without external current applied, $\delta\omega$ = 0.23 deg. This is close to and a little higher than the experimental resolution given by the angular divergence of the beam. If we increase the transport current, but staying below the critical current value, absolutely no change is observed. When the applied current is higher than the critical current of 2.5 A, a slight increase of the rocking curve width $\Delta\omega$ is observed. The reason is that the transport current I imposed by the external generator has to fulfill the Maxwell-Ampere equation. As the moving Bragg planes are observed invariant by translation, one can neglect the in-plane field gradient and the Maxwell-Ampere equation reduces to $\mu_0 J$= curl B which physically represents a curvature of the field lines over the thickness of the sample. This bending is responsible of the slight increase of the rocking curve width that we observed and gives a good idea of the transport current value that is flowing in the bulk of the sample ($I_{bulk}$ = {2W B}\$\mu_0$ $\Delta\omega$). The variation of the rocking curve width as function of I is shown in fig. 5. One can estimate that no bulk current is present for I<Ic and that a bulk current that proportional to (I-Ic) for I>Ic. It is very interesting to modify the critical current by sanding the surface and to observe that indeed, the current always flows in the sample bulk only above Ic. The critical current can be decreases back to small values by etching by a mixing of acetic acid and $H_2O_2$. A non-homogeneous sample can also be prepared by inhomogeneous surface treatment. In this case, it should be pointed out that the current is not flowing parallel to the surface anymore. For this reason, some vortex loops appear perpendicular to the current which can be at the origin of many very strange effects often observed in high Tc materials (figure 6).

We have performed the same kind of experiments in crystals of $NbSe_2$ in order to compare the different states of FLL responsible for the anomalous transport properties. We first tried to observed the simplest case of the ZFC FLL. This latter is supposed to reflect the ordered and equilibrium state, because it corresponds to the state with a low critical current. The diffraction patterns exhibit an ordered hexagonal lattice. We performed also ω rocking curves. Small widths are obtained by analyzing the peaks with Lorentzian fits (Fig. 7). We obtain, for the ZFC FLL and without external current applied, $\delta\omega = 0.232$ deg. This is close to and a little higher than the experimental resolution given by the angular divergence of the beam. If we increase the transport current, but staying below the critical current value, absolutely no change is observed. When the applied current is higher than the critical current of 2.5 A, a slight increase of the rocking curve width $\Delta\omega$ is observed. It is clear that the points are separated by values just slightly higher than the experimental resolution (given mainly by the mechanical precision of the angle during the rotation) . One can nevertheless estimate that no bulk current is present for I<Ic and that a bulk current that is rather worth (I-Ic) is observed for I>Ic. This result, added to the moderate critical current measured and to the V(I) curve that exhibits the usual form $R_{ff}$(I-Ic) and not as $R_{ff}$ I indicates that under critical superficial current due to surface pinning and bulk dissipative over critical current offer a natural explanation for this behavior. Concerning SANS coupled with transport experiments in $NbSe_2$, it is quite natural to cite Yaron et al experiments /14/, whom purpose was to measure the longitudinal correlation length characteristic of FLL order. Yaron et al observed a narrowing of the rocking curve that they attributed to an improvement of FLL order. On the contrary, we observe here, what was previously observed in other superconductors that the rocking curve broadens as the over critical current penetrates the bulk. As this is a simple consequence of the Maxwell equations, it appears not clear to us why such effect was not observed by Yaron et al. A possible interpretation is that the rocking curve reported are made in the direction perpendicular to those reported here. In such case and as observed in Nb-Ta samples, a very small narrowing can be observed but it can reasonably rather be attributed to a smoothing of FLL Bragg planes spacing (due to the homogeneous bending its the perpendicular direction) than to a change in a correlation length.

The FLL in $NbSe_2$, created by a ZFC, appears so quite similar to the FLL in conventional type II superconductor with moderate critical current. In fact, more differences are expected after a FC, because in this case the V(I) curve looks very peculiar. The samples we used for the SANS experiments are larger than those usually employed for transport properties, it is thus important to precise that we have measured V(I) curves (Fig.8) very similar to what was already studied in details by others. They exhibit a hysteretic behavior, with a S shape for the first run after FC and a linear and reversible behavior for all ramps of current after. It is worth noting that such V(I) curves classically observed in $NbSe_2$ are a particular case of vortex dynamic. For a large number of type II superconductors, these effects are not observed.

If the ZFC FLL observed in Fe doped $NbSe_2$ was close to what we can call a conventional FLL, obtaining information on the FLL structure after FC was a real difficulty. For the same Bragg conditions as for the ZFC state, we do not see any scattered intensity. The first idea was that the FLL was so strongly disordered that the Bragg peaks were considerably broadened and thus almost invisible. But this is not the right reason, as evidenced in Fig.8 where the corresponding rocking curve is shown. Compared with the results of Fig. 7, it is easy to see that the Bragg conditions for FLL has changed and that the rocking curve exhibits a double peak, what is quite unusual. The sum of the integrated intensity contained in these two peaks compared well, within error bars, to the integrated intensity of the Bragg peak of the ZFC FLL, and the widths of the peaks are comparable too. Consequently, we can not attribute those strange Bragg peaks to a FLL disorder in its proper sense. It would better correspond to two very similar families of FLL that are ordered, but slightly tilted from the magnetic field direction from few tens of degrees. We can eliminate a rotation coming from Doppler shift because the FLL frame is not moving. Another possibility is that we are now observing two FLL possessing two different Bragg planes spacing because of different magnetic densities. But this assumption would imply a field gradient of more than half the magnetic field present in the sample. This looks hardly compatible with the strong interaction between the flux lines that limits the compressibility of the vortex array.

Finally, it is quite reasonable to think that the two peaks we observe is the signature of the "two phases" observed by Marchevski et al using scanning hall ac probe /5/. Their experiments show that two states possessing different critical current are coexisting in the region of the peak effect. Our SANS experiment offers complementary information. The fact that the two peaks are very similar is not in favor of two states that are characterized by different bulk underlying disorder. The shift between these two peaks indicates that the two FLL are slightly tilted by static and small in plane field components. It looks clear that these field components are due to a peculiar and non-symmetric distribution of screening current. Following previous authors, we adopt the point of view that the border of the sample is a region of a high superficial current density. Both Bragg peaks cover roughly the same surface and we can speculate that the width of the sample is divided into two part of roughly the same dimension, i.e. 3 mm for each. We know that the low critical current is 2.5 A and that it corresponds to a superficial value of 2 A/cm. At the same time, we have measured, when the peak effect is at its maximum, a ratio of about 7. With the reasonable assumption that it corresponds to a state where the high critical current state invades most of the sample, we can deduced that the high critical value is about 14 A/cm. Using the Ampere theorem and making a superposition with the top and bottom surfaces, we find that two sheets transporting currents generates bulk components of magnetic field which are worth 2.5 and 18 G respectively. This is not so far from the measured values (9 and 24 G), considering the highly schematic picture which is used here.

In this picture, the FC state is made by large loops of current, that are as many non dissipative paths for a transport current. Increasing the transport current induces a preferential direction and one loop should be turned off. This implies that one of the two tilted FFL disappears. If the transport current is increased again up to the high critical current, the second loop disappear. All the flux lines are now along the main magnetic field and the Bragg angle returns close to its normal position. Finally, the surfaces can not transport more non-dissipative current, the excess penetrates the bulk and the flow becomes resistive. The V(I) curve returns to a classical behavior in the linear form.

## Conclusion

In conclusion, using a model of pinning by surface irregularities in anisotropic superconductors, we have developed a calculation of the critical current which allows estimating quantitatively the critical current in both the high critical current phase and in the low critical current phase. The only adjustable parameter of this model is the surface roughness which can be measured by AFM and modified by FIBE processes. The agreement between the measurements and the model is really very impressive. In this framework, the peak effect is due to co-existence of two different vortex phases with different critical currents. A detailed discussion of the vortex geometry is necessary to understand the details of the peak effect. Neutron diffraction data in $NbSe_2$ crystals in presence of transport current support this point of view.

[1] see for example: T. Giamarchi , S. Bhattacharya, Vortex Phases in ``High Magnetic Fields: Applications in Condensed Matter Physics and Spectroscopy", C. Berthier et al., 9, 314, Springer-Verlag (2002).
[2] Y.Simon and P.Thorel,Phys.Lett.35A,450 (1971) and P.Thorel,Y.Simon and A.Guetta, J.Low Temp.Phys.11,333 (1973).
[3] P.Mathieu and Y.Simon,Europhys.Lett.5,67 (1988).
[4] Y. Paltiel, D. T. Fuchs, E. Zeldov, Y. N. Myasoedov, H. Shtrikman, M. L. Rappaport, and E. Y. Andrei, Phys. Rev. B 58, R14763-R14766 (1998).
[5] M. Marchevsky, M.J. Higgins, and S. Bhattacharya, Nature (London) 409, 591 (2001) and Phys. Rev. Lett. 88, 087002 (2002).
[6] Y. Fasano, M. Menghini, F. de la Cruz, Y. Paltiel, Y. Myasoedov, E. Zeldov, M. J. Higgins, and S. Bhattacharya, Phys. Rev. B 66, 020512 (2002)
[7] A. Pautrat, J. Scola, C. Goupil, Ch. Simon, C. Villard, B. Domengès, Y. Simon, C. Guilpin, and L. Méchin, Phys. Rev. B 69, 224504 (2004)
[8] G.Lazard, P.Mathieu, B.Plaçais, J.Mosqueira, Y.Simon, C.Guilpin and G.Vacquier, Phys.Rev.B 65, 064518 (2002).
[9] B.Plaçais, P.Mathieu and Y.Simon, Solid State Commun.71,177 (1989).
[10] Y.Simon, B.Pla çais and P.Mathieu, Phys.Rev. B 50,3503 (1994).
[11] A.Pautrat, Ch. Simon, J. Scola, C. Goupil, A. Ruyter, L. Ammor, P. Thopart, D. Plessis, to be published in Europhysics Journal B.
[12] J. Schelten, H. Ullmaier and G. Lippmann, Phys. Rev. B {\bf 12},1772 (1975).
[13] A. Pautrat, C. Goupil, Ch. Simon, D. Charalambous, E. M. Forgan, G. Lazard, P. Mathieu, and A. Brûlet, Phys. Rev. Lett. 90, 087002 (2003).
[14] U. Yaron, P.L. Gammel, D.A. Huse, R.N. Kleiman, C.S.Oglesby, E. Bucher, B. Batlogg, D. Bishop, K. Mortensen, K. Clausen, C.A. Bolle and F. De La Cruz, Phys. Rev. Lett. 73, 2748 (1994).

FIG.1: The critical current of the virgin niobium microbridge as function of the magnetic field at 4.2K for two diffrent surface states. In the inset is shown the calculated value of the angle $\alpha$ for the two surface roughnesses. The calculation of $\alpha$ was performed using the value of $\varepsilon$ shown in the bottom part of the figure (calculated).

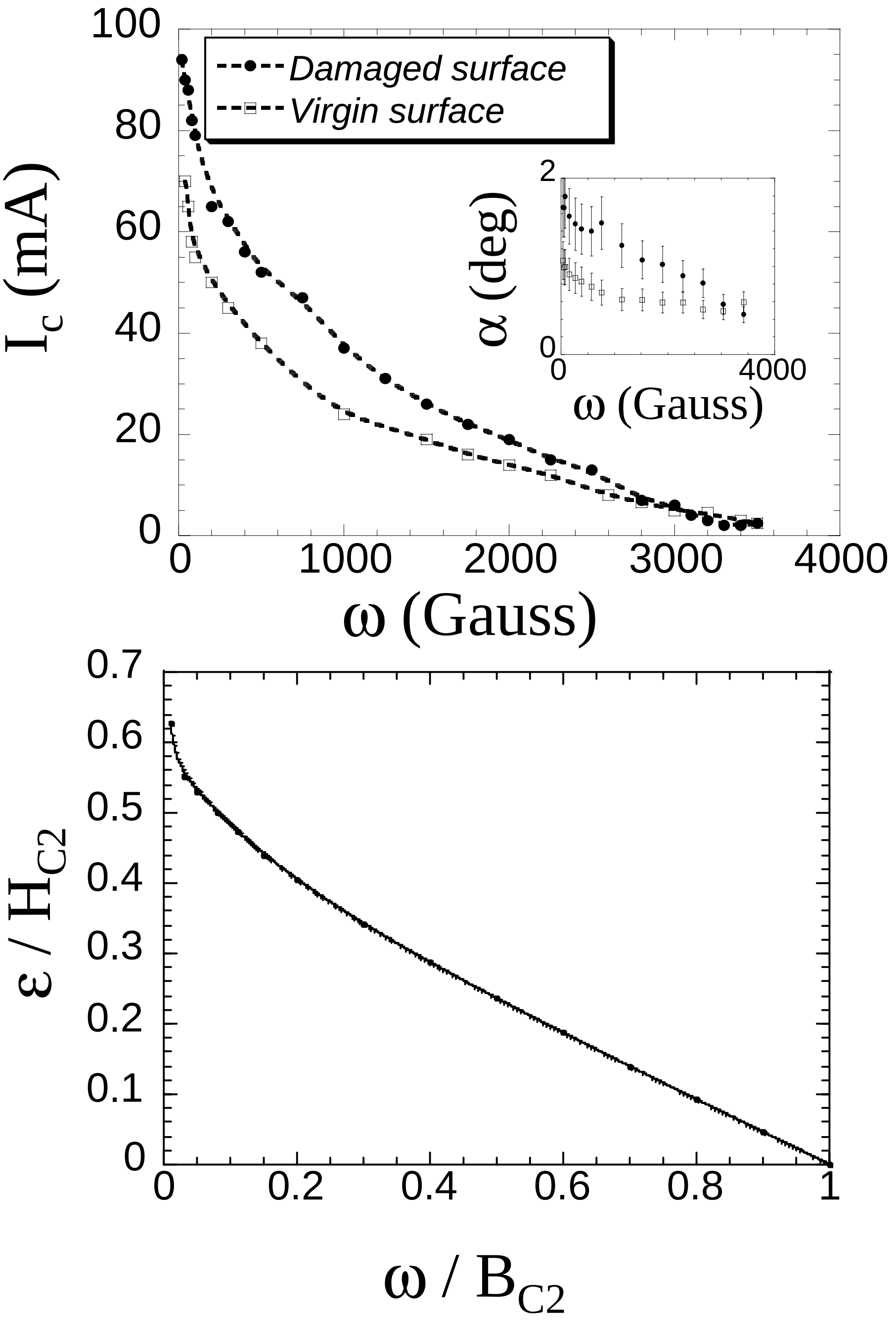

FIG 1

FIG.2: The critical current as function of the magnetic field at 4.2K for the same $NbSe_2$ sample before and after sandblasting, showing the influence of the surface roughness on the peak effect.

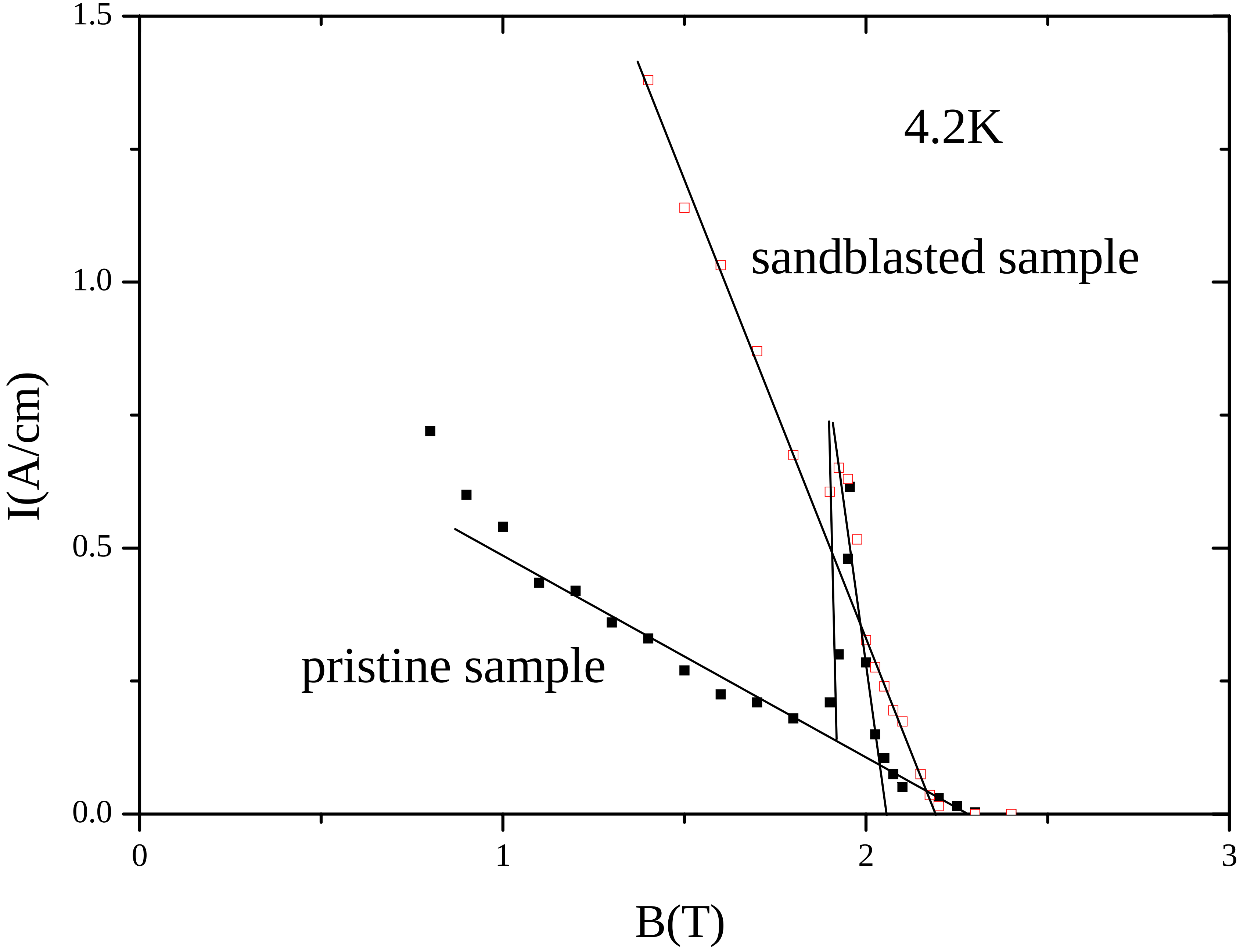

Fig 2

FIG.3: Typical IV curves of a Bi-2212 microbridge under different magnetic fields (top curve: above the peak effect from T to T, bottom part: at 300G)

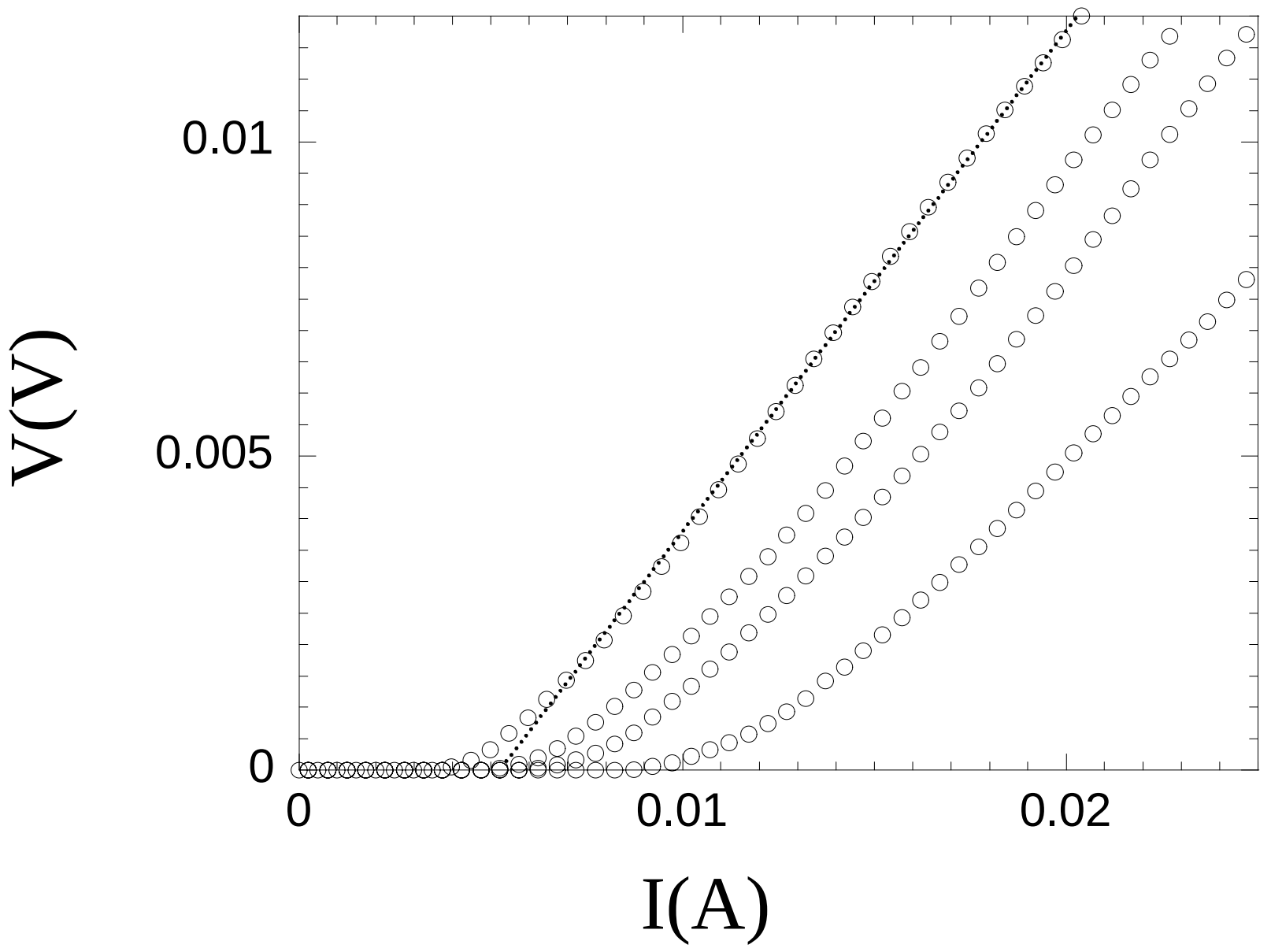

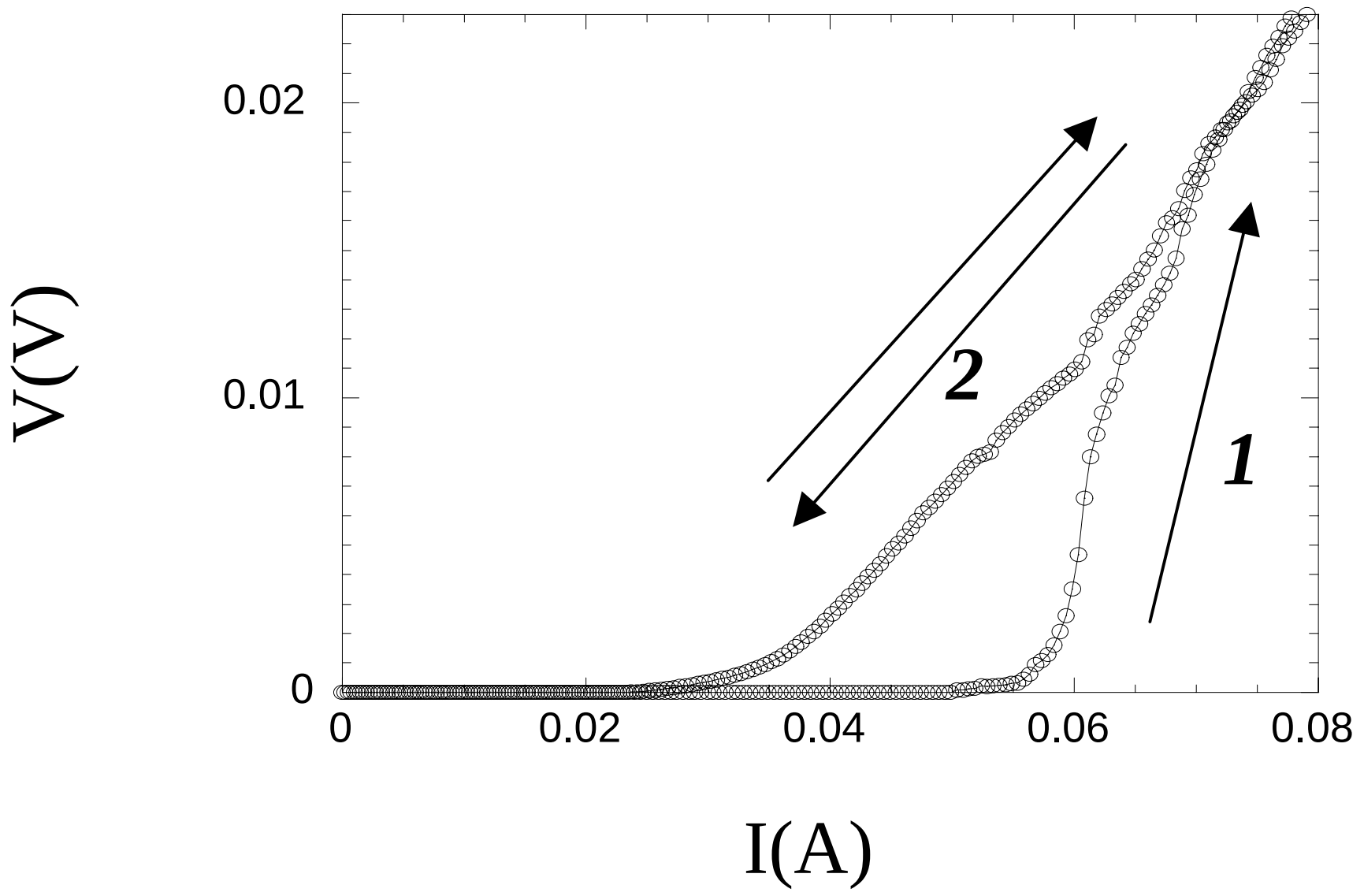

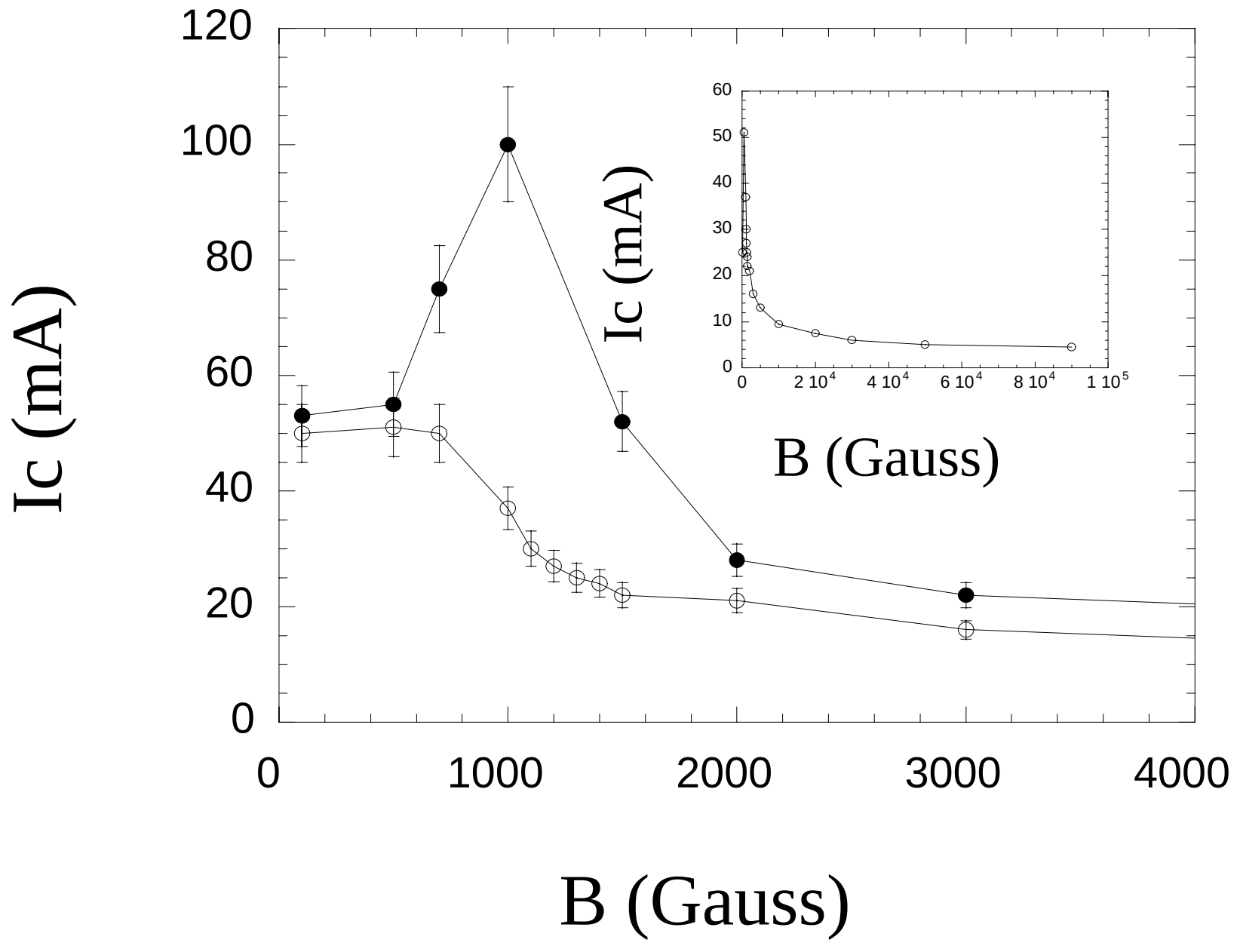

FIG.4: The critical current as function of the magnetic field at 4.2K for the Bi-2212 microbridge. The two different curves correspond to the small and the high critical currents observed in the bottom part of figure 3 respectiveley. In the inset, the data are presented for higher magnetic field, showing the coherence with the proposed model.

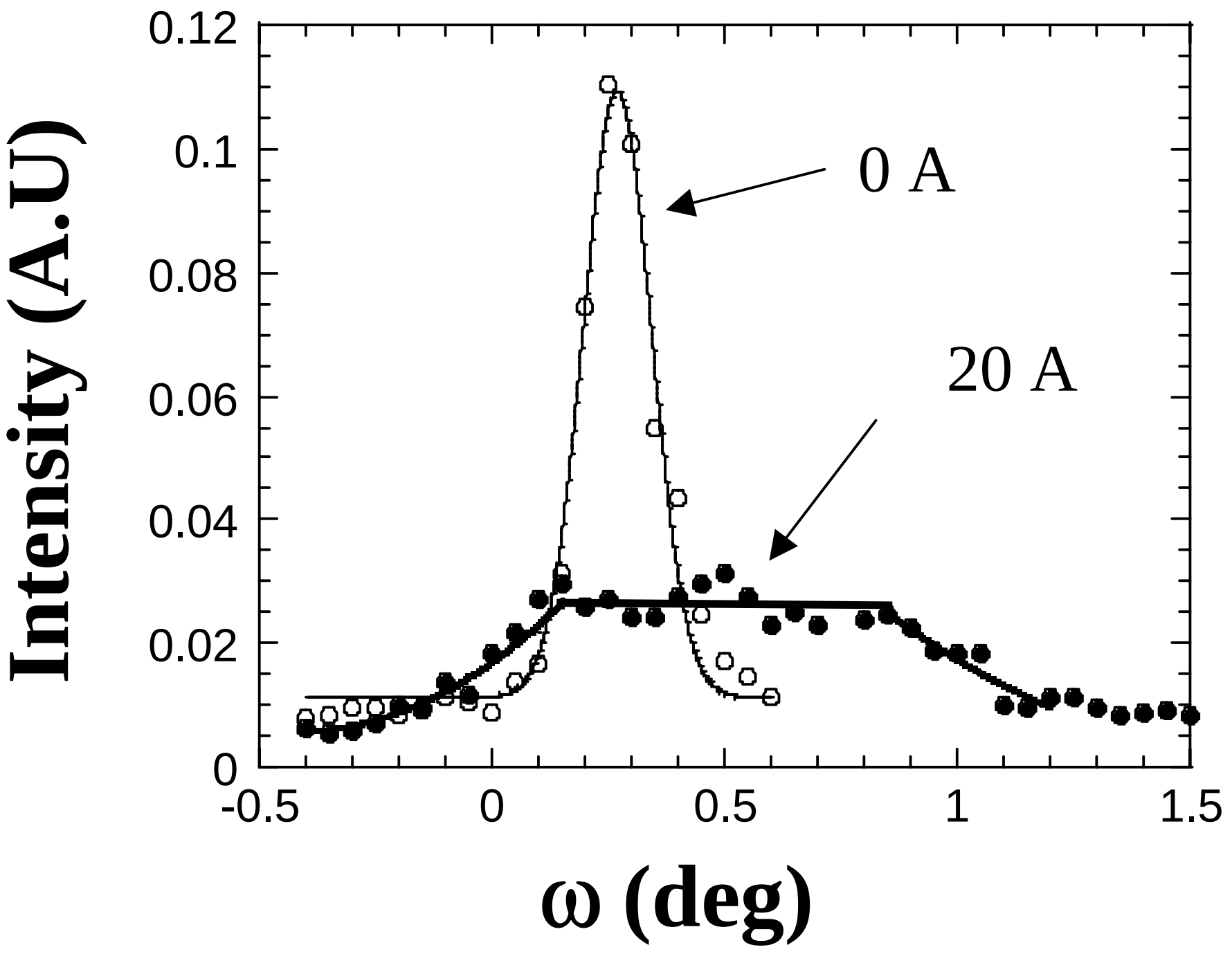

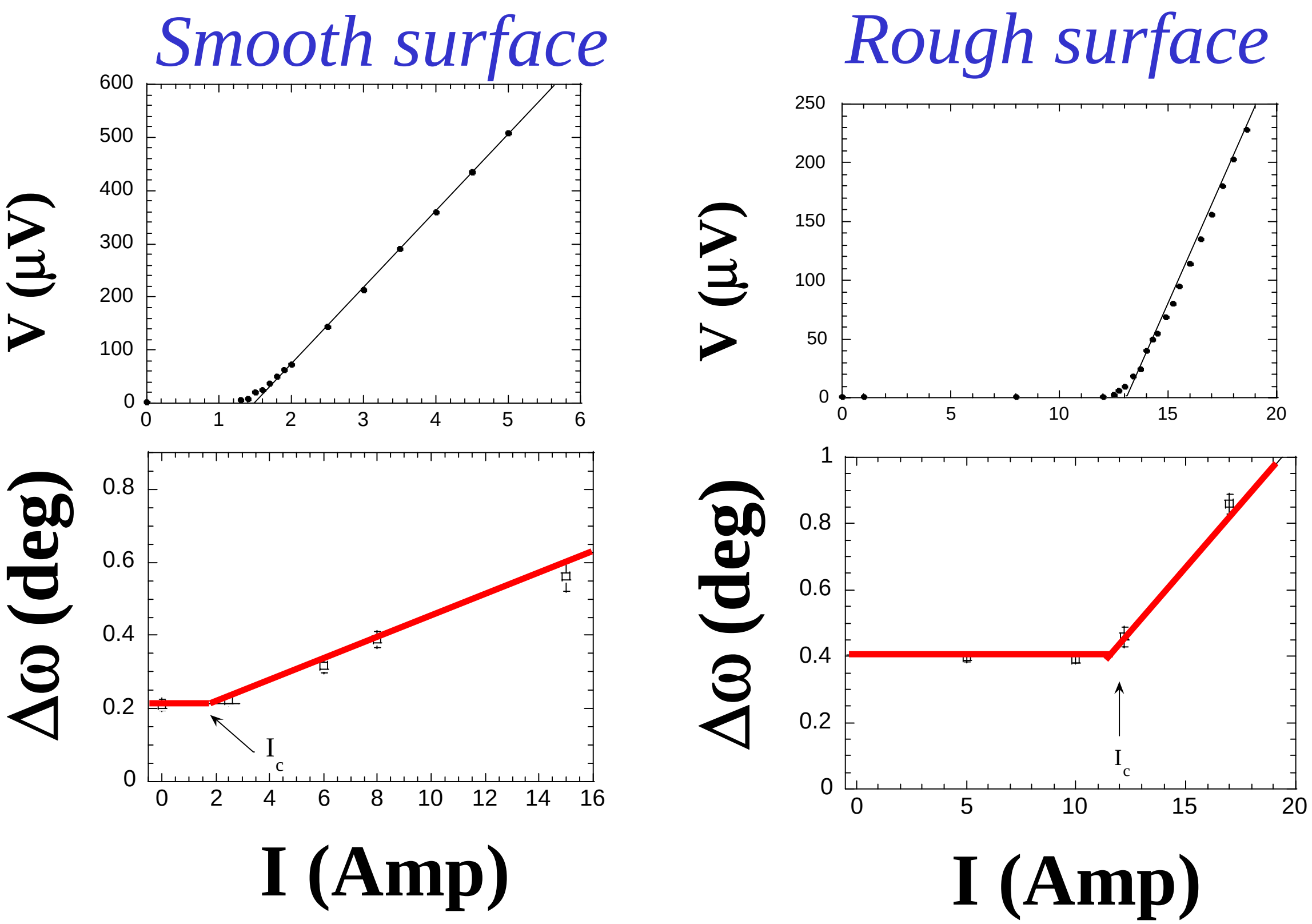

FIG.5: The rocking curves at different currents at 4.2K in a Pb-In sample. In the bottom part, the IV curves are presented, compared to the broadening of the rocking curve, for two different suface roughness of the same sample.

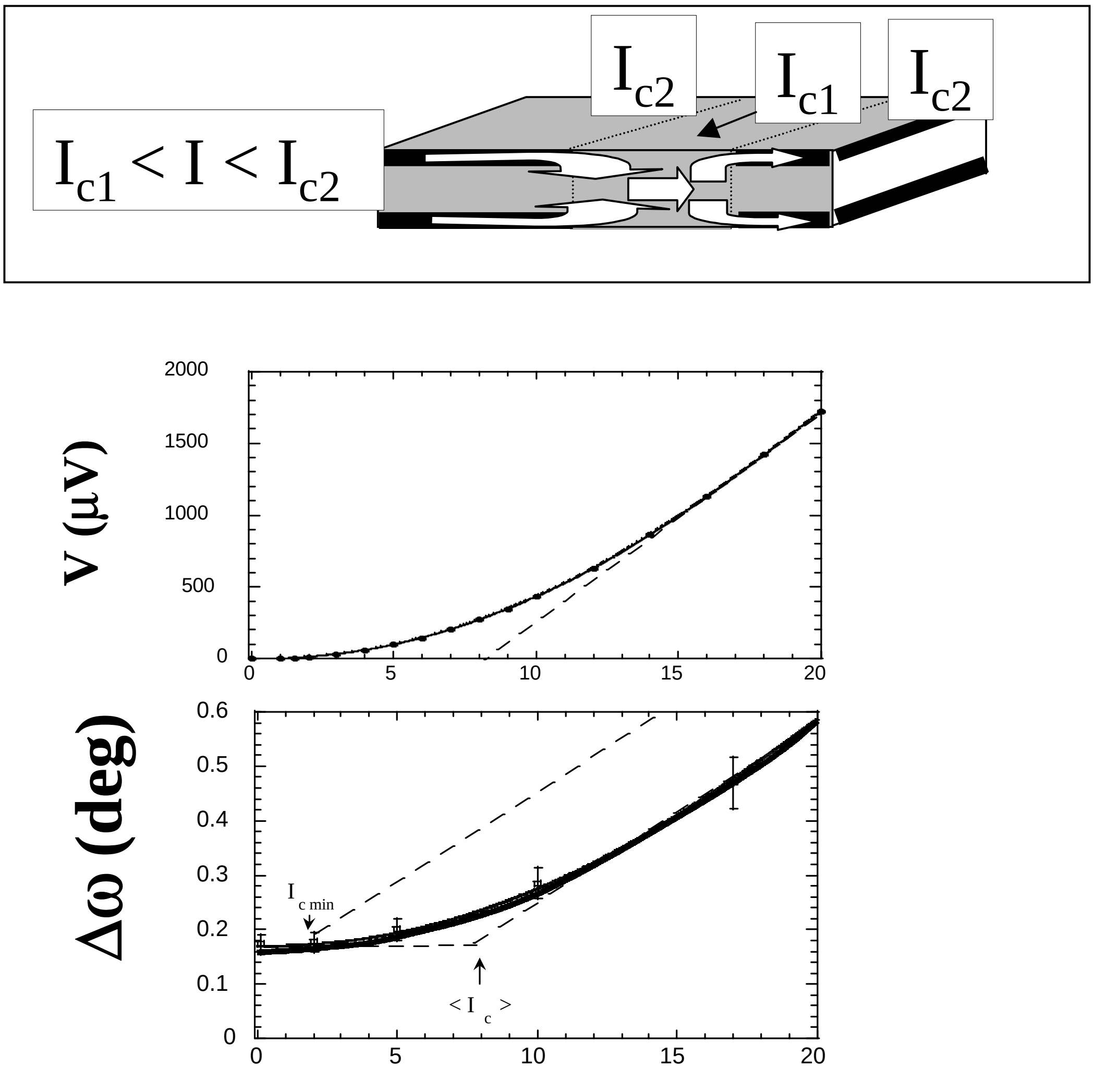

FIG.6: The comparison of the IV curve and the broadening of the rocking curve in a sample in which the critical current is not homogeneous. The solid line of the bottom part of the picture is the calculation obtained assuming the critical current inhomogeneity measured in the upper part of the picture on the IV curve. A schematic drawing of the current flow is also shown.

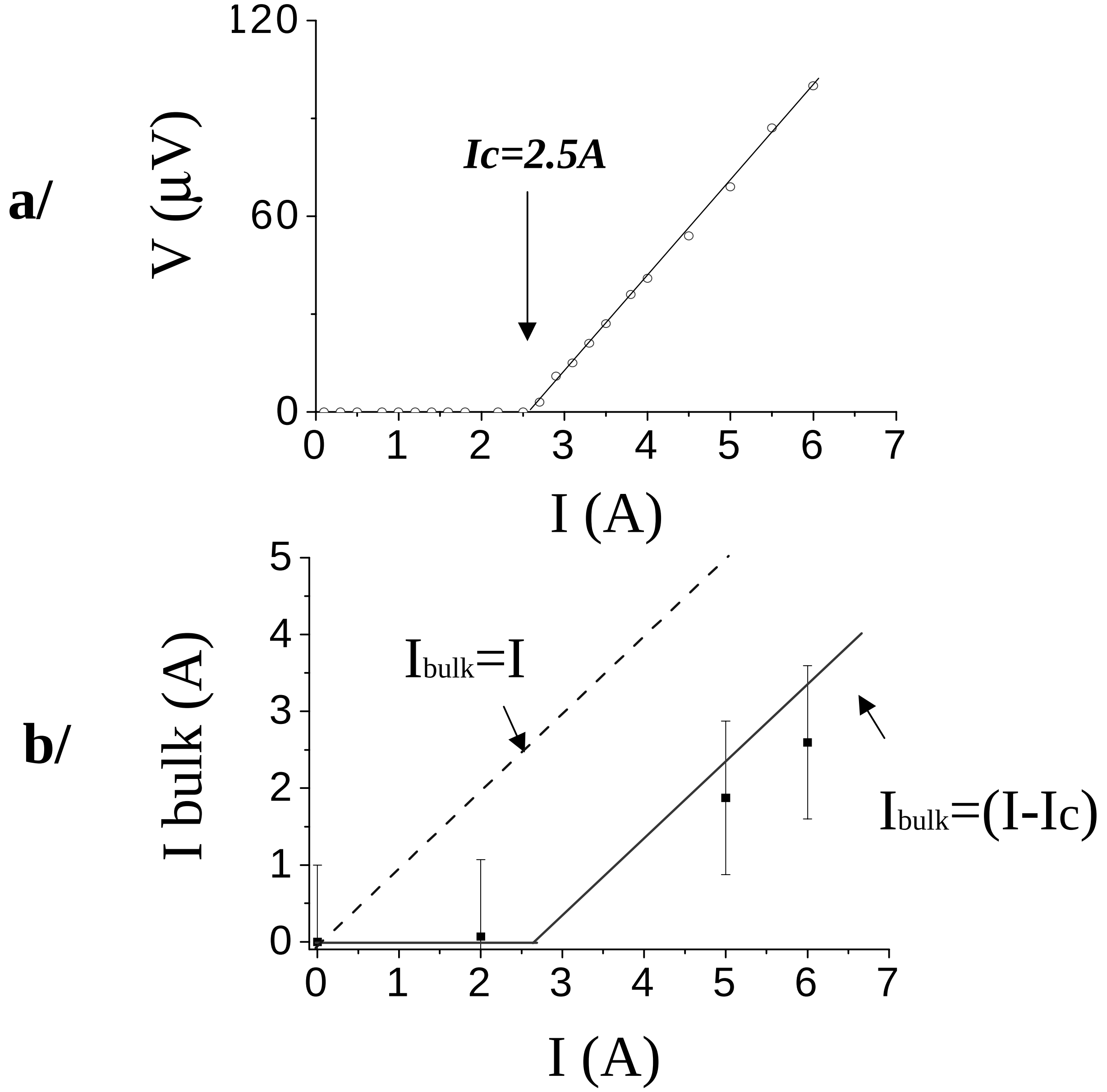

F

FIG 7: The comparison of the IV curve and the broadening of the rocking curve in a $NbSe_2$ sample at 2.5K and 3000G. The solid line of the bottom part of the picture is the calculation obtained assuming the critical current is flowing at the sample surface. The dotted line corresponds to the assuption of the current flowing homogeneously in the sample bulk.

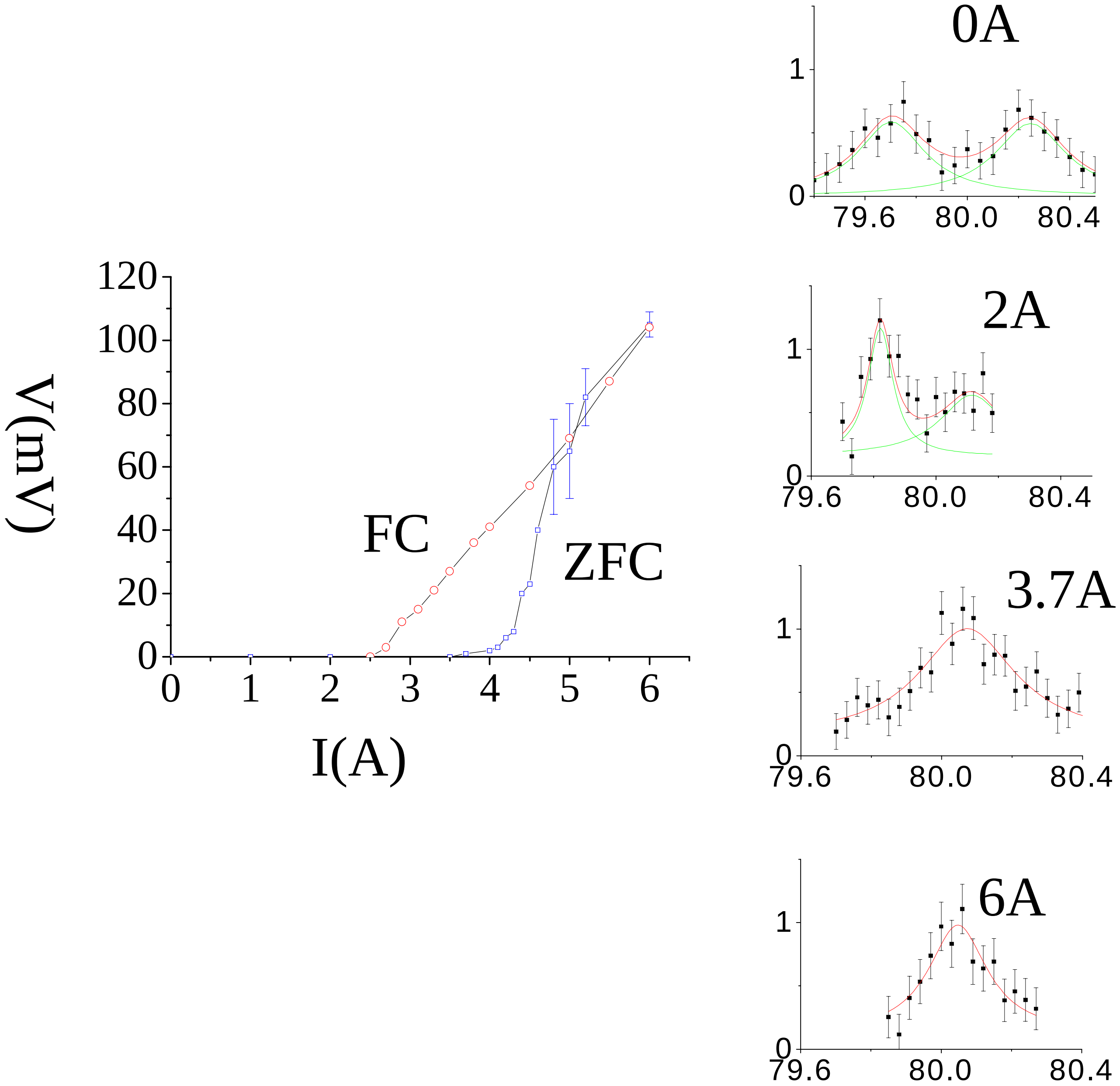

FIG 8: The rocking curves obtained in the field cooled procedure. On the right part of the figure, the corresponding IV curves are shown.